\documentclass[journal]{IEEEtran}
\ifCLASSINFOpdf
\else
   \usepackage[dvips]{graphicx}
\fi
\usepackage{url}
\usepackage{cite}
\usepackage[caption=false,font=normalsize,labelfont=sf,textfont=sf]{subfig}
\usepackage{graphicx}
\usepackage{amsmath}
\usepackage{amsfonts}
\usepackage{array}
\usepackage{mathtools}
\usepackage{multirow}

\newcolumntype{C}[1]{>{\centering\let\newline\\\arraybackslash\hspace{0pt}}m{#1}}
\newcommand{\specialcell}[2][c]{\begin{tabular}[#1]{@{}c@{}}#2\end{tabular}}

\begin{document}

\title{HILCodec: High-Fidelity and Lightweight Neural Audio Codec}

\author{Sunghwan Ahn, Beom Jun Woo, Min Hyun Han, Chanyeong Moon, and Nam Soo Kim, \IEEEmembership{Senior Member, IEEE}
\thanks{This work was supported by Institute of Information \& communications Technology Planning \& Evaluation (IITP) grant funded by the Korea government(MSIT) (No.2021-0-00456, Development of Ultra-high Speech Quality  Technology for Remote Multi-speaker Conference System)}
\thanks{Sunghwan Ahn, Beom Jun Woo, Min Hyun Han, Chanyeong Moon, and Nam Soo Kim are with the Department of Electrical and Computer Engineering and the Institute of New Media and Communications, Seoul National University, Seoul 08826, South Korea (e-mail: shahn@hi.snu.ac.kr; bjwoo@hi.snu.ac.kr; mhhan@hi.snu.ac.kr; cymoon@hi.snu.ac.kr; nkim@snu.ac.kr).}
}

\markboth{}
{Shell \MakeLowercase{\textit{et al.}}: Bare Demo of IEEEtran.cls for IEEE Journals}
\maketitle

\begin{abstract}
The recent advancement of end-to-end neural audio codecs enables compressing audio at very low bitrates while reconstructing the output audio with high fidelity. Nonetheless, such improvements often come at the cost of increased model complexity. In this paper, we identify and address the problems of existing neural audio codecs. We show that the performance of the SEANet-based codec does not increase consistently as the network depth increases. We analyze the root cause of such a phenomenon and suggest a variance-constrained design. Also, we reveal various distortions in previous waveform domain discriminators and propose a novel distortion-free discriminator. The resulting model, \textit{HILCodec}, is a real-time streaming audio codec that demonstrates state-of-the-art quality across various bitrates and audio types.
\end{abstract}

\begin{IEEEkeywords}
Acoustic signal processing, audio coding, codecs, generative adversarial networks, residual neural networks
\end{IEEEkeywords}

\IEEEpeerreviewmaketitle

\section{Introduction}
\IEEEPARstart{A}{n audio} codec is a system that compresses and decompresses audio data. It comprises an encoder which analyzes the input audio, a quantizer, and a decoder that reconstructs the output audio. Audio codecs have several objectives \cite{coding-book}. First, they aim to compress the input using the minimum number of bits. Second, they strive to preserve the perceptual quality of the output, making it as similar as possible to the original input. Third, they seek to maintain a low algorithmic delay for streaming applications. Lastly, they are designed to have low computational complexity so that they can be used on a variety of devices, including mobile and embedded systems.

Traditional audio codecs leverage the expertise of speech processing, audio signal processing, and psychoacoustics to achieve their objectives. For speech coding, based on the source-filter model of speech production \cite{lp}, linear predictive coding (LPC) \cite{lpc} is widely adopted because of its high compression rate. Instead of compressing the waveform directly, LPC represents the input speech as an autoregressive model and quantizes the model's coefficients. These coefficients are then dequantized and used to synthesize the output speech. However, the limitation of LPC in capturing high-frequency components leads to the use of transform coding \cite{tc, amrwbp, uni-speech-audio, aspec, aac}, particularly when preserving high-frequency components is crucial. This method first transforms the input waveform into another domain, such as sub-bands \cite{mp1}, wavelet coefficients \cite{wavelet}, or a time-frequency spectrogram \cite{mp2}, to obtain less correlated components. It then allocates available bits so that quantization noise for each component is psychoacoustically well-distributed \cite{psycho}. Finally, the decoder applies an inverse transform to obtain the output waveform.

Recently, the remarkable advancements of machine learning have spurred active research into audio coding based on deep neural networks (DNNs). One research direction involves using a DNN as a post-processing enhancement module for existing codecs to suppress coding artifacts at low bitrates \cite{codec-enhance1, PostGAN, codec-enhance2}. In some studies, a DNN decoder is trained on top of a traditional human-designed encoder. The input for such neural decoder can be the output representations of an encoder after quantization \cite{lpcnet, wavenet-codec}, or the raw bitstream itself \cite{neur-bit-dec}. Another promising area is an end-to-end neural audio codec \cite{mlnn, e2esc, soundstream, encodec, descript, audiodec, hificodec}, where instead of relying on manually designed encoders and quantizers, the entire encoder-quantizer-decoder framework is trained in a data-driven way. Leveraging the power of generative adversarial networks (GAN) \cite {GAN}, end-to-end neural audio codecs have demonstrated state-of-the-art audio quality under low bitrate conditions. On the other hand, these improvements in audio quality are often attributed to an increase in network size, leading to substantial computational complexity. Furthermore, some of these codecs are not streamable. While there exist neural audio codecs that can be processed in real time streaming manner, it results in compromised performance. %Given the growing demand and potential for neural audio codecs in real-world services, the development of a codec that is lightweight, streamable, and maintains high fidelity presents a timely and valuable opportunity.

%In this paper, we challenge the common practice of simply augmenting model complexity. Our main focus is on dissecting and addressing the issues discovered from the existing generators and discriminators which hinder model performance. Combined with a lightweight backbone network, we propose \textit{HILCodec}, an end-to-end neural audio codec that satisfies the aforementioned requirements. It is capable of compressing diverse types of audio data at bitrates ranging from 1.5 kbps to 9.0 kbps, with a sampling rate of 24kHz. \textit{HILCodec} can operate in real-time on a single thread of a CPU while demonstrating comparable or superior audio quality to other traditional and neural audio codecs.

In this paper, we aim to develop a neural audio codec model with the following properties:
\begin{itemize}
    \item \textbf{Streaming}: When an input audio is divided into short chunks and provided in a streaming manner, the model must be able to encode and decode each chunk sequentially.
    \item \textbf{Lightweight}: The number of model parameters should be small.
    \item \textbf{Low-complexity}: The number of multiply-accumulate (MAC) opeartions should be low.
    \item \textbf{High-fidelity}: The codec’s output should be perceptually similar to the input across various bitrates and audio types.
\end{itemize}

To fulfill streaming, lightweight, and low-complexity capabilities, we designed a lightweight version of the widely used SEANet architecture \cite{seanet} (Section \ref{sec:generator}). For high fidelity, our strategy is three-fold. First, we incorporate $L_2$-normalization (Section \ref{sec:l2norm}) and multi-resolution spectrogram inputs (Section \ref{sec:specblock}), which enhance the model's performance and training stability with a minimal increase in computational complexity. Second, we identify an issue with the SEANet architecture where an increase in network depth does not always lead to an improvement in audio quality and may even result in quality degradation (Section \ref{sec:problem-of-seanet}). This problem has not been addressed explicitly in previous studies on audio coding. Through theoretical and empirical analysis, we deduce that this discrepancy is due to the exponential increase in variance of activations during the forward pass. To mitigate this, we propose a variance-constrained design, which effectively scales performance in line with network depth (Section \ref{sec:VCRB} to Section \ref{sec:zero-init}). Lastly, we show that existing waveform domain discriminators often distort input signals before feeding them into neural networks (Section \ref{sec:problem-of-disc}). As a solution, we introduce a distortion-free discriminator that utilizes multiple filter banks to enhance perceptual quality (Section \ref{sec:mfbd}). When combined with a time-frequency domain discriminator, it surpasses the performance of conventional discriminators in GAN-based audio codecs in terms of subjective quality. These innovative approaches significantly contribute to the overall effectiveness of our proposed model, \textit{HILCodec}.

Our main contributions are summarized as follow:
\begin{itemize}
    \item We propose a lightweight generator backbone architecture that runs on devices with limited resources.
    \item We identify a problem associated with the SEANet generator: increasing network depth does not necessarily lead to improved performance. We address this by adopting a variance-constrained design.
    \item Based on signal processing theories, we propose a distortion-free multi-filter bank discriminator (MFBD).
    \item We compare \textit{HILCodec} against the state-of-the-art traditional codec and neural audio codecs across various bitrates and audio types. The evaluation results demonstrate that our model matches or even surpasses the performance of previous codecs, even with reduced computational complexity.
\end{itemize}

Audio samples, code, and pre-trained weights are publicly available\footnote{https://github.com/aask1357/hilcodec}.

\section{Related Works}
\subsection{Neural Vocoder}
A vocoder is a system that synthesizes human voice from latent representations. The study on vocoders originated from \cite{dudley} and has since been central to digital speech coding. Recently, neural vocoders have brought significant attention due to their applicability to text-to-speech (TTS) systems. A neural vocoder synthesizes speech from a given mel-spectrogram, while a separate neural network generates the mel-spectrogram from text. Various neural vocoders have been proposed based on deep generative models, including autoregressive models \cite{wavenet, par-wavenet, wavernn}, flow \cite{waveglow}, diffusion \cite{diffwave, wavegrad}, and others. Among these, vocoders based on the GAN framework have been popular due to their high-quality output and fast inference time. MelGAN \cite{MelGAN} introduced the Multi-Scale Discriminator (MSD) which operates on a waveform after average pooling, and a feature matching loss to stabilize adversarial training. Parallel WaveGAN \cite{par-wavegan} introduced a multi-resolution short-time Fourier transform (STFT) loss, which outperforms the single-resolution counterpart. HiFi-GAN \cite{hifigan} proposed a well-designed generator and the Multi-Period Discriminator (MPD) which operates on downsampled waveform. Avocodo \cite{avocodo} aimed to reduce aliasing in the generator and waveform domain discriminators by introducing the Cooperative Multi-Band Discriminator (CoMBD) and the Sub-Band Discriminator (SBD). Some models used time-frequency domain discriminators \cite{univnet, bigvgan}. Due to the similarity of the tasks, developments in the field of neural vocoders have been exploited for various neural audio codecs. In this work, we also utilize GAN-based training techniques from the line of this trend, while analyzing signal distortion incurred by existing discriminators and proposing a novel filter bank-based distortion-free discriminator. This approach aligns with a method proposed in \cite{avocodo}. Nonetheless, compared to previous works, our discriminator can handle distortions around the transition bands and shows superior audio quality.

\subsection{End-to-end Neural Audio Codec}
Soundstream \cite{soundstream} is a GAN-based end-to-end streaming neural audio codec trained on general audio. By utilizing residual vector quantizer \cite{rvq1, rvq2}, it facilitates the training of a single model that supports multiple bitrates. EnCodec \cite{encodec} introduced a novel loss balancer to stabilize training. AudioDec \cite{audiodec} proposed an efficient GAN training scheme and modified the model architecture of HiFi-GAN. Despite decent quality for speech coding, it requires 4 threads of a powerful workstation CPU for real-time operation. HiFi-Codec \cite{hificodec} and Descript Audio Codec \cite{descript} also adopted the generator and discriminators from HiFi-GAN, each showing state-of-the-art performance for clean speech and general audio compression. However, they are not streamable and their computational complexity is high. FreqCodec \cite{funcodec} proposed a lightweight frequency domain model with low computational complexity, showing exceptional performance in objective evaluations. However, it did not undergo a subjective evaluation. Penguins \cite{penguins} and LightCodec \cite{lightcodec} employed low-complexity neural coding modules, each demonstrating subjective quality superior to EnCodec. Nonetheless, their evaluations were limited to a speech corpus. Additionally, these models are not publicly available, and their capabilities for streaming remain unspecified. In this paper, we introduce a model architecture that is streamable, lightweight, and has low computational complexity. Furthermore, we pinpoint and address a previously unexplored problem within the waveform domain generator, proposing a solution that enhances audio quality beyond that of prior models over various audio types.

\section{Generator}
\label{sec:generator}

\begin{figure*}
    \centering
    \includegraphics[width=1.0\linewidth]{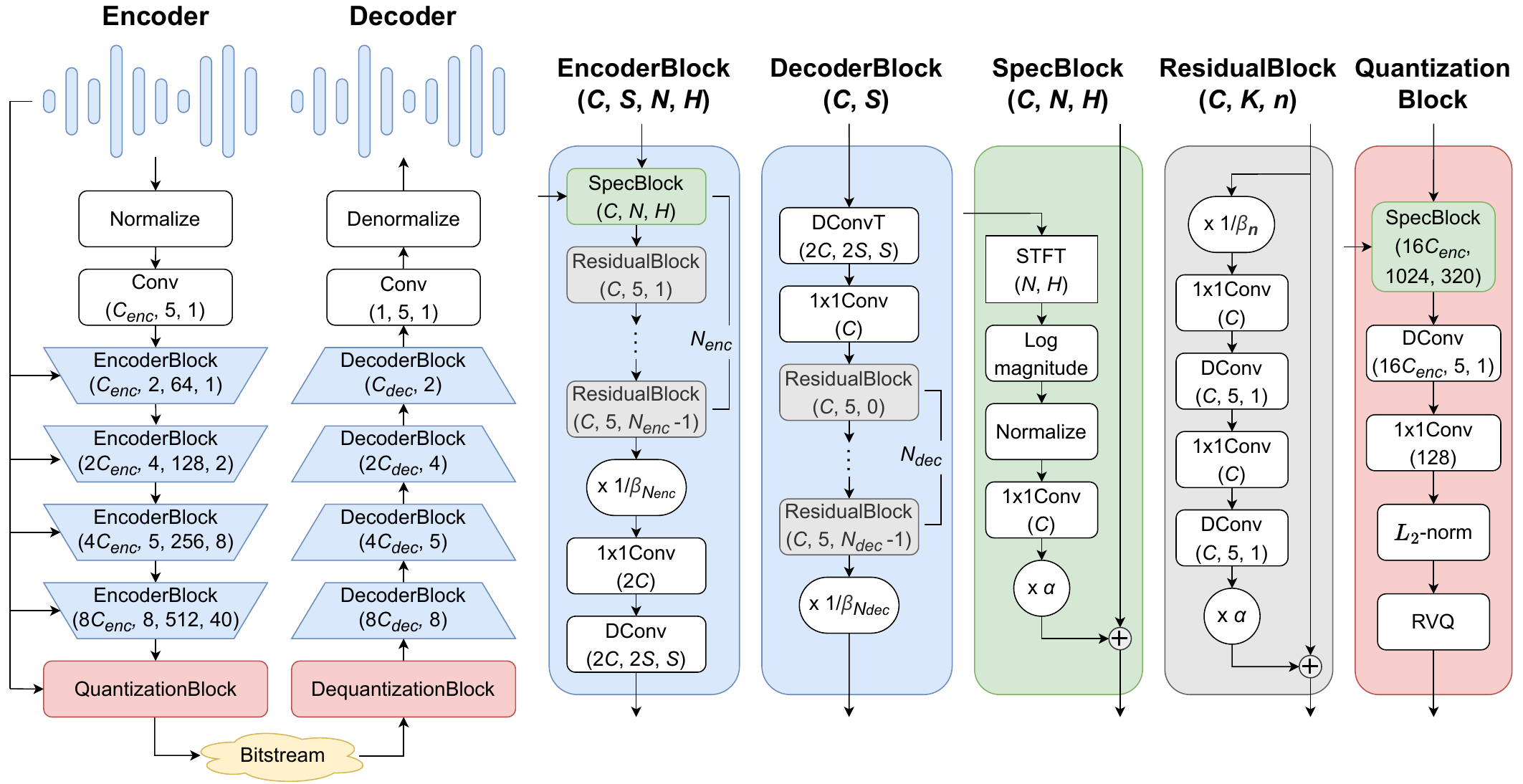}
    \caption{\textit{HILCodec} model architecture. Conv(\textit{C}, \textit{K}, \textit{S}), DConv(\textit{C}, \textit{K}, \textit{S}), and DConvT(\textit{C}, \textit{K}, \textit{S}) represent a convolution, a depthwise convolution, and a depthwise transposed convolution respectively, each with output channels of \textit{C}, a kernel size of \textit{K}, and a stride of \textit{S}. 1x1Conv(\textit{C}) denotes a pointwise convolution with output channels of \textit{C}.}
    \label{fig:model-arch}
\end{figure*}

\textit{HILCodec} is an end-to-end neural audio codec that leverages the time-domain fully convolutional layers. The overall model architecture is illustrated in Fig. \ref{fig:model-arch}. Following \cite{soundstream}, \textit{HilCodec} is composed of an encoder, a residual vector quantizer (RVQ), and a decoder (Section \ref{sec:enc-dec}, \ref{sec:rvq}). Given a raw waveform input, the encoder generates a feature vector for every 320 samples. The  decoder reconstructs the waveform from a sequence of quantized features. It is designed in favor of four desired properties: streaming, lightweight, low-complexity, and high-fidelity. To ensure the streaming property, all convolutions are causal \cite{str-sea}. To achieve lightweight and low-complexity properties, we utilize depthwise separable convolutions \cite{dws}. To achieve high-fidelity without significantly increasing the model complexity, we introduce $L_2$-normalization (Section \ref{sec:l2norm}), spectrogram blocks (Section \ref{sec:specblock}), and a variance-constrained design (Section \ref{sec:vcd}). In the following subsections, we provide a detailed explanation of each part of the model.

\subsection{Encoder-Decoder}
\label{sec:enc-dec}
Similar to \cite{seanet}, the encoder is composed of a convolutional layer with $C_{enc}$ output channels followed by 4 encoder blocks and a quantization block. Each encoder block contains a spectrogram block (Section \ref{sec:specblock}) and $N_{enc}-1$ residual blocks, followed by a strided convolution for downsampling. The strides are set to 2, 4, 5, and 8. The quantization block is composed of a spectrogram block, two convolutions, an $L_2$-normalization layer (Section \ref{sec:l2norm}), and an RVQ with codebook vectors of $128$ dimensions.

The decoder mirrors the encoder's structure, replacing strided convolutions with transposed convolutions for upsampling, and substituting the quantization block with a dequantization block. The dequantization block is composed of a dequantization layer, a pointwise convolution, and a depthwise convolution. Each decoder block contains $N_{dec}$ residual blocks. The final convolution has $C_{dec}$ input channels. An ELU activation function \cite{elu} is inserted before every convolution, except the first one of the encoder. A hyperbolic tangent activation is inserted after the last convolution of the decoder to ensure that the final output waveform lies within the range of $-1$ and $1$. Following \cite{encodec}, we apply weight normalization \cite{wn} to every convolution to further enhance the model's performance.

In the domain of neural image compression, the prevalent approach involves the use of a compact encoder coupled with a larger decoder \cite{img-comp1, img-comp2}. Various neural audio codecs also adopted this paradigm \cite{audiodec, descript}. It is reported that downsizing the encoder has a minimal effect on performance, while a reduction in the decoder size significantly degrades audio quality \cite{soundstream}. The architecture of \textit{HILCodec} is designed in accordance with these approaches. The encoder is configured with $C_{enc}$ set to 64 and the number of residual blocks $N_{enc}-1$ set to $2$, while the decoder is set up with $C_{dec}=96$ and $N_{dec}=3$.

\subsection{Residual Vector Quantizer}
\label{sec:rvq}
We employ the same training mechanism for the RVQ as in \cite{soundstream}. The codebook vectors in the RVQ are initialized using k-means clustering. As proposed in \cite{VQVAE}, when the encoder's output $O_e$ is quantized to $O_q$ by the RVQ, the gradient $\nabla_{O_q}L$ is passed directly to $O_e$ during backpropagation. To ensure proximity between $O_e$ and $O_q$, a mean squared error loss between $O_e$ and $O_q$ is enforced on the encoder. The codebook vectors are updated using an exponential moving average (ema) with a smoothing factor of $0.99$. To prevent some codebook vectors from being unused, we reinitialize codebook vectors whose ema of usage per mini-batch falls below $0.5$.

\subsection{$L_2$-normalization}
\label{sec:l2norm}
Since the advent of VQVAE \cite{VQVAE}, there has been a continuous exploration for superior vector quantization. One such method is the $L_2$-normalization of the encoder outputs and codebook vectors, which has been proven to improve the codebook utilization and the final output quality of a generative model \cite{image-ivqgan, descript}. In our experiments, although $L_2$-normalization did not consistently improve audio quality across all audio types, we applied it to our model because it stabilizes mixed-precision training \cite{fp16} (Section \ref{sec:ablation-generator}). Unlike \cite{image-ivqgan} and \cite{descript}, we do not directly apply $L_2$-normalization to the codebook vectors. As we employ RVQ, we naturally apply the $L_2$-normalization only to the encoder output before quantization.

\begin{figure*}
    \centering
    \includegraphics[width=1.0\linewidth]{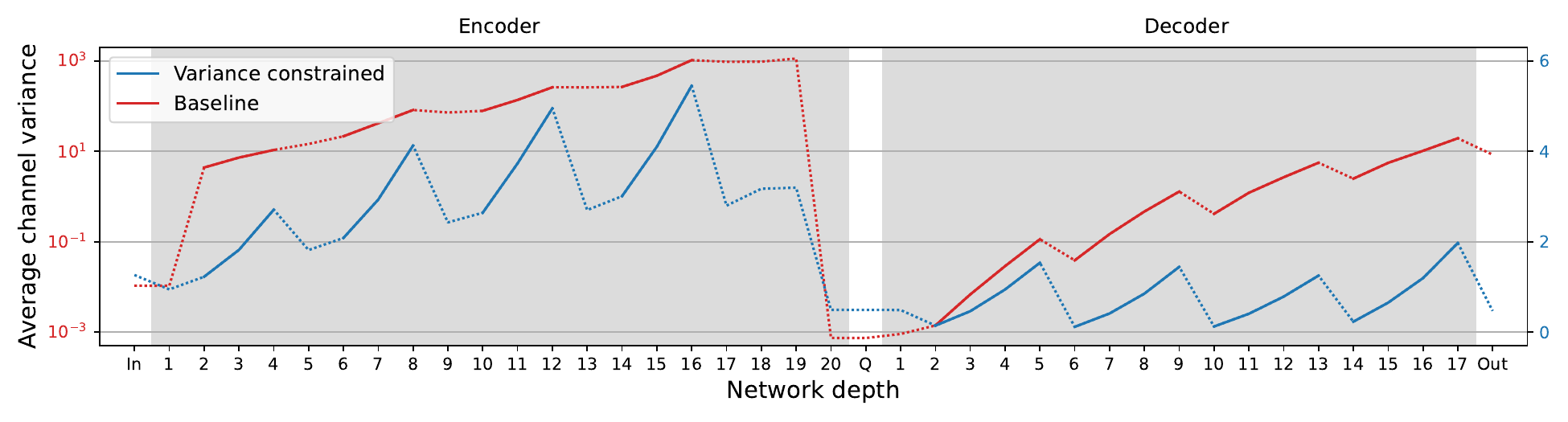}
    \caption{Comparison of the average channel variance after initialization according to network depth. On the x-axis, each number represents the network depth of the encoder and the decoder, and the terms “In”, “Q”, and “Out” denote the input waveform, the output of the residual vector quantizer, and the output waveform, respectively. The logaritmic y-scale on the left corresponds to the baseline model, while the linear y-scale on the right corresponds to the variance constrained model. The solid lines represent the residual blocks.}
    \label{fig:avg-ch-var}
\end{figure*}

\subsection{Spectrogram Block}
\label{sec:specblock}
Traditional audio codecs exploit human-designed feature extractors for compression. In contrast, an end-to-end neural codec learns a feature extractor in a data-driven manner without requiring any expertise on audio coding. While a larger model can learn better features, it comes at the cost of increased computational complexity. To circumvent this, we extract additional features akin to those used in transform coding and feed them into the encoder network. Specifically, we insert a spectrogram block at the beginning of each encoder block and  quantization block. As shown in Fig. \ref{fig:model-arch}, a spectrogram block is composed of a log-magnitude spectrogram extractor, a normalization layer (Section \ref{sec:inout_norm}), a pointwise convolution, and a skip connection. We provide log-magnitude spectrograms with hop sizes of $\{1, 2, 8, 40, 320\}$ and Fourier transform sizes of $\{64, 128, 256, 512, 1024\}$. The hop sizes are set so that the time resolution of each spectrogram matches that of the input of each encoder block and quantization block. These multi-resolution spectrograms assist the encoder with a limited parameter size in learning richer features. Importantly, we preserve the streaming nature of our model by using a causal STFT to obtain the spectrogram.

\subsection{Variance-Constrained Design}
\label{sec:vcd}
For deep neural networks, proper signal propagation in the early stages of training is particularly important. However, we have identified that the existing model architecture exhibits undesired signal propagation immediately after initialization (Section \ref{sec:problem-of-seanet}). As a solution, we propose a variance-constrained design (Sections \ref{sec:VCRB} to Section \ref{sec:zero-init}). This approach enhances audio quality without increasing model complexity (Section \ref{sec:ablation-generator}).
\subsubsection{Variance Explosion Problem of SEANet}
\label{sec:problem-of-seanet}
The SEANet architecture, which forms the backbone of the \textit{HilCodec} generator, is a widely adopted model for end-to-end neural audio codecs \cite{soundstream, encodec, descript, hificodec, audiodec}. However, we observed that the audio quality does not improve consistently as the depth of the model increases (Fig. \ref{fig:visqol}). This observation aligns with the results reported in \cite[Table A.3]{encodec}, where the objective score for audio quality did not increase even after the number of residual blocks in the encoder and decoder was tripled. If the dataset size is limited, the performance of a DNN model may not improve as the model size increases. However, since our model is trained on a large audio corpus (Section \ref{sec:dataset-and-training}), it is natural to expect that the performance scales with the model size. To investigate this discrepancy, we concentrate on the signal propagation within the model's forward pass.

Consider a tensor $\mathbf{x}$ with a shape of $B \times C \times T$ where $B$ denotes the batch size, $C$ denotes the channel size, and $T$ denotes time length. For simplicity, we define variance, or $\mathbf{Var}(\mathbf{x})$, as the average channel variance of $\mathbf{x}$, computed by taking the variance across the batch and time dimensions, and then averaging across the channel dimension:
\begin{multline}
   \mathbf{Var}(\mathbf{x}) \coloneqq \frac{1}{C}\sum_{c} \Bigg[ \frac{1}{BT} \sum_{b}\sum_{t} \mathbf{x}^2_{b,c,t} \\ - \bigg( \frac{1}{BT} \sum_{b}\sum_{t} \mathbf{x}_{b,c,t} \bigg)^2 \Bigg].
\end{multline}
% \begin{equation}
%    \mathbf{Var}(\mathbf{x}) \coloneqq \frac{1}{C}\sum_{c} \Bigg[ \frac{1}{BT} \sum_{b}\sum_{t} \mathbf{x}^2_{b,c,t} - \bigg( \frac{1}{BT} \sum_{b}\sum_{t} \mathbf{x}_{b,c,t} \bigg)^2 \Bigg].
% \end{equation}
Fig. \ref{fig:avg-ch-var} illustrates the signal propagation plot \cite{nfnet1} of \textit{HILCodec} right after initialization. Note that the y-axis of the baseline model is on a logarithmic scale ranging from $10^{-3}$ to $10^3$. In the encoder, the variance increases exponentially until it reaches the final layer, the $L_2$-normalization layer. In the decoder, the exponential growth of the variance persists until the end. The reasons and downsides for such a variance explosion phenomenon can be well understood in the literature on deep normalized networks \cite{skipinit, nfnet1, fixup}.

In the field of computer vision, the combination of batch normalization \cite{BN} and skip connection \cite{resnet} showed great performance gain and enabled very deep networks to be stably trained. Theoretical analyses have been conducted on the signal propagation of these networks to explain their success. Consider a network solely composed of residual blocks. The $n$-th residual block takes the form $\mathbf{x}_{n+1}=\mathbf{x}_n+f_n(\mathbf{x}_n)$, where $\mathbf{x}_n$ is the input and $\mathbf{x}_{n+1}$ is the output. We refer to $f_n$ as the residual branch. Right after initialization, given that the input and the model weights are independent, $\mathbf{Var}(\mathbf{x}_{n+1})=\mathbf{Var}(\mathbf{x}_n)+\mathbf{Var}(f_n(\mathbf{x}_n))$. Reference \cite{skipinit} showed that if $f_n$ is a properly initialized network without any normalization layer, $\mathbf{Var}(f(\mathbf{x}_n))\approx \mathbf{Var}(\mathbf{x}_n)$, leading to $\mathbf{Var}(\mathbf{x}_{n+1})\approx 2\mathbf{Var}(\mathbf{x}_{n})$. This implies that the variance increases exponentially according to the network depth $n$. On the other hand, if $f_n$ is a normalized network with appropriate initialization, $\mathbf{Var}(f_n(\mathbf{x}_n))\approx 1$, resulting in $\mathbf{Var}(\mathbf{x}_{n+1})\approx n+1$. This linear growth is beneficial to the model's signal propagation $-$ it implies that the output of the residual branch $f_n(\mathbf{x}_n)$ contributes only a $\frac{1}{n+1}$ fraction to $\mathbf{x}_{n+1}$. As a result, the output of each residual block is dominated by a skip connection, leading to stable training even for very deep networks. On the other hand, in unnormalized networks, both the residual branch and skip connection contribute equally to the output of each residual block, hindering signal propagation at the beginning of training.

Back to our model, since the SEANet architecture doesn't contain any normalization layer, it experiences an exponential growth in variance as a function of network depth, resulting in performance degradation as illustrated in Fig. \ref{fig:visqol}. Also, due to the high dynamic range of the baseline model's signal propagation, mixed-precision training fails when the number of residual blocks in each encoder/decoder block exceeds two. Prior works have indirectly addressed this issue by increasing the network's width instead of its depth. For instance, HiFi-GAN employs multi-receptive field fusion modules, where the outputs from multiple parallel residual blocks are aggregated to yield the final output, instead of sequentially connecting each block. Furthermore, AudioDec extends this structure by explicitly broadening the network width using grouped convolutions.

Our goal is to directly solve the problem of exponential variance growth that occurs immediately after initialization, rather than merely circumventing it. The most straightforward solution is to incorporate a normalization layer into every residual block. Nevertheless, our preliminary experiments with various normalization layers, including batch normalization and layer normalization \cite{LN}, resulted in even poorer performance. We can only conjecture the reasons $-$ perhaps the batch size was insufficient due to hardware constraints, or the normalization layers we used were ill-suited for the waveform domain. To address this, instead of relying on normalization layers, we propose a variance-constrained design (VCD) which facilitates proper signal propagation. This encompasses the residual block design, input and output normalization, and zero initialization of residual branches. As shown in Fig. \ref{fig:visqol}, after applying our methods, the audio quality improves according to the network depth. Note that the proposed design is not constrained to the SEANet architecture $-$ it can be applied to various models that have skip connections.

\subsubsection{Variance-Constrained Residual Block}
\label{sec:VCRB}
In this subsection, we present the Variance-Constrained Residual Block (VCRB) designed to downscale the variance of the output of each residual branch such that the variance increases linearly with the network depth right after initialization. It is motivated by the NFNet block \cite{nfnet1} in image domain. A VCRB can be mathematically written as
\begin{equation}
\label{nfnet}
\mathbf{x}_{n+1}=\mathbf{x}_n+ \alpha \cdot f_n\bigg(\frac{\mathbf{x}_n}{\beta_n}\bigg),
\end{equation}
where $\alpha$ and $\beta_n$ are fixed scalars to be discussed. We consider the residual branch $f_n$ right after initialization. Also, we assume that the network input $\mathbf{x}_0$ consists of {\it i.i.d.} random variables with a mean of $0$ and a standard deviation of $1$, and $f_n(\cdot)$ is a sample function of a random process indexed by $\mathbf{x}$. There are three key components in \eqref{nfnet}. Firstly, we carefully initialize the parameters of $f_n$ such that $\mathbf{Var}(f_n(\mathbf{x}_n))=\mathbf{Var}(\mathbf{x}_n)$ (details in Appendix \ref{appendix-variance}). Secondly, we analytically calculate $\mathbf{Var}(\mathbf{x}_n)$ as in \eqref{resunit-var}, and set
\begin{equation}
    \beta_n=\sqrt{\mathbf{Var}(\mathbf{x}_n)}.
\end{equation}
This ensures that the variance of the input and the output of $f_n(\cdot)$ is $1$:
\begin{equation}
    \mathbf{Var}\Bigg(f_n\bigg(\frac{\mathbf{x}_n}{\beta_n}\bigg)\Bigg)=\mathbf{Var}\bigg(\frac{\mathbf{x}_n}{\beta_n}\bigg)=1.
\end{equation}
Thirdly, we set 
\begin{equation}
    \alpha=\frac{1}{\sqrt{N}},
\end{equation}
where $N$ is the number of residual blocks in an encoder or decoder block. Since $\mathbf{Var}(\mathbf{x}_0)=1$, the variance of the output from the $n$-th residual block can be calculated as follows:
\begin{equation}
\label{resunit-var}
\begin{split}
    \mathbf{Var}(\mathbf{x}_n) & =\mathbf{Var}(\mathbf{x}_{n-1})+\mathbf{Var}\Bigg(\alpha \cdot f_{n-1}\bigg(\frac{\mathbf{x}_{n-1}}{\beta_{n-1}}\bigg)\Bigg) \\
    &=\mathbf{Var}(\mathbf{x}_{n-1})+\frac{1}{N} \\
    &=\cdots \\
    &=\mathbf{Var}(\mathbf{x}_0)+\frac{n}{N} \\
    &=1+\frac{n}{N}.
\end{split}
\end{equation}
\eqref{resunit-var} implies that the encoder/decoder blocks composed of VCRB possess the desired property of linear variance increment along the network depth.

\subsubsection{Other Building Blocks}
For the spectrogram blocks described in Section \ref{sec:specblock}, we treat them as a special type of residual block. This means that similar to the other residual blocks, we scale the output of each spectrogram block $g(\cdot)$ as:
\begin{equation}
\label{specblock-var}
    \mathbf{y} = \mathbf{x} + \alpha \cdot g(\mathbf{s}),
\end{equation}
where $\mathbf{s}$ is a spectrogram input. Assuming $\mathbf{s}$ has unit variance, \eqref{specblock-var} ensures $\mathbf{Var}(\mathbf{y})=1+1/N$.

As illustrated in Fig. \ref{fig:model-arch}, at the end of each encoder block, there exists a strided convolution for downsampling. Since it follows $N_{enc}$ sequential residual blocks, its input variance equals $2$ according to \eqref{resunit-var}. Therefore, we divide its input by $\beta_{N_{enc}}=\sqrt{2}$. Similarly, we divide the output of each decoder block by $\beta_{N_{dec}}=\sqrt{2}$.

\subsubsection{Input/Output Normalization}
\label{sec:inout_norm}
We further normalize the input and output of both the encoder and decoder. We compute the mean and standard deviation of $10000$ randomly sampled audio chunks from the training dataset. At the beginning of the encoder, we insert a normalization layer using pre-calculated statistics. We also insert normalization layers for multi-resolution spectrogram inputs. In the decoder, right after the final convolution and prior to the hyperbolic tangent activation, we insert a de-normalization layer. The final element we normalize is the output of the encoder (equivalently the input of the decoder before quantization), denoted as $\mathbf{z}$. Since we apply $l_2$-normalization before quantization as discussed in Section \ref{sec:l2norm}, the mean square of $\mathbf{z}$ is $\mathbb{E}(\mathbf{z}^2)=1/C$, where $C$ is the channel dimension of $\mathbf{z}$. Ignoring the square of the mean of $\mathbf{z}$, we multiply $\sqrt{C}$ to $\mathbf{z}$ after $l_2$-normalization so that the variance is approximately $1$ as given by
\begin{equation}
    \mathbf{Var}(\sqrt{C} \cdot \mathbf{z})
    = C \cdot \mathbb{E}(\mathbf{z}^2) - C \cdot \mathbb{E}(\mathbf{z})^2
    \approx C\cdot \mathbb{E}(\mathbf{z}^2)=1.
\end{equation}
After applying variance-constrained blocks and input/output normalization, the variance of the network right after initialization remains in a reasonable dynamic range as illustrated in Fig. \ref{fig:avg-ch-var}.

\subsubsection{Zero-Initialized Residual Branch}
\label{sec:zero-init}
For deep residual networks, it has been empirically shown that initializing every residual block to be an identity function at the beginning of training can further enhance the performance \cite{bag-of-tricks, deep-vit}. Following \cite{nfnet1, rezero}, we add a scalar parameter initialized to zero at the end of each residual branch, which we also found beneficial for the fast convergence of the model. While this zero-initialization scheme results in a modification of $\beta_n$ in Section \ref{sec:VCRB}, we found that ignoring this effect does not detrimentally impact the performance of our model.

\section{Discriminator}

\begin{figure*}
    \centering
    \includegraphics[width=1.0\linewidth]{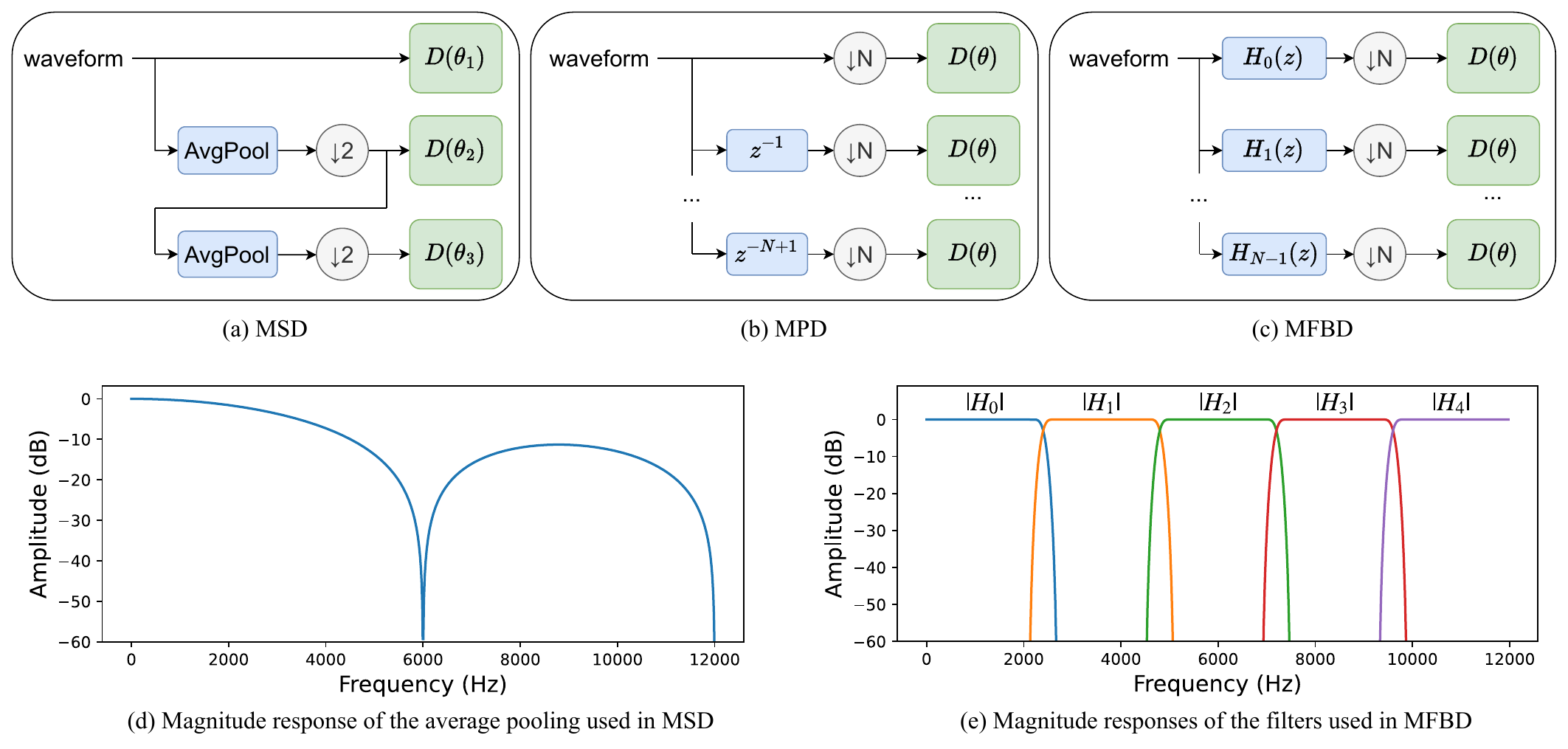}
    \caption{(a)-(c): Comparison of different discriminators. (d)-(e): Magnitude responses of filters they use. (a) AvgPool denotes an average pooling. (b) $z^{-n}$ denotes $n$-sample delay. (a)-(c) $\downarrow$$N$ denotes downsampling by a factor of $N$, and $D(\theta)$ denotes a sub-discriminator with a parameter $\theta$. MSD uses distinct sub-discriminators for different hierarchies while MPD and MFBD use shared parameters across the sub-bands.}
    \label{fig:disc}
\end{figure*}

We apply two types of discriminators: a novel aliasing-free waveform domain discriminator and a widely used time-frequency domain discriminator.
\subsection{Multi-Filter Bank Discriminator}
\subsubsection{Problems of previous discriminators}
\label{sec:problem-of-disc}
Digital audio encompasses a significantly large number of samples within a short time span. For instance, an audio clip that is 3 seconds long at a sampling rate of 24kHz contains 72,000 samples. Consequently, waveform domain discriminators typically accept audio inputs after downsampling. As shown in Fig. \ref{fig:disc}(a) and (b), the Multi-Scale Discriminator (MSD) \cite{MelGAN} uses average pooling, and the Multi-Period Discriminator (MPD) \cite{hifigan} relies on a plain downsampling. The combination of MSD and MPD has achieved significant success in the field of neural vocoders and has been employed in various neural audio codecs \cite{hificodec, audiodec, descript}. However, from a signal processing perspective, these discriminators induce distortion to the input signal. MPD relies on a plain downsampling, which causes aliasing of the signal. In the case of MSD, average pooling can be interpreted as a low-pass filter using a rectangular window. Fig. \ref{fig:disc}(d) depicts the magnitude response of a 4-tap average pooling used in MSD. The passband is not flat, the transition band is too wide and not steep, and the attenuation level in the stopband is too low. It is evident that average pooling distorts input signal and is insufficient to prevent aliasing caused by subsequent downsampling. Furthermore, after average pooling, MSD is limited to modeling low-frequency components only. This highlights the need for a more sophisticated waveform domain discriminator capable of preventing distortion and modeling the entire frequency spectrum.

\subsubsection{Proposed discriminator}
\label{sec:mfbd}
We propose a novel Multi-Filter Bank Discriminator (MFBD). As illustrated in Fig. \ref{fig:disc}(c), it initially transforms the waveform into $N$ sub-bands using a filter bank, followed by a downsampling with a factor of $N$. Each sub-band is then fed into a single discriminator with shared weights. This enables the discriminator to observe the entire frequency components with minimum distortion. We used pseudo-quadratic mirror filters (PQMF) \cite{pqmf} as the filter bank. Similar to \cite{hifigan}, MFBD is composed of multiple sub-discriminators with the number of sub-bands $N \in \{1, 2, 3, 5, 7, 11\}$. This approach has two beneficial effects. Sub-discriminators with small $N$ can model short-term dependencies, while those with large $N$ can model long-term dependencies. Additionally, because we use a coprime number of sub-bands, even the distortion in transition bands can be covered by other sub-discriminators. The model architecture of MFBD is configured in the same way as MPD.

A similar approach was also introduced in Avocodo \cite{avocodo}. It uses PQMF to obtain sub-band features and feeds the entire features to a discriminator. The key differences are that MFBD uses a shared discriminator across different sub-bands, while Avocodo uses distinct discriminators for each sub-band. We also utilize multiple sub-discriminators with a co-prime number of sub-bands to cope with distortion in transition bands. Our experiments demonstrate the superiority of MFBD over the discriminators used in Avocodo (Section \ref{sec:ablation-disc}). For more detailed comparisons, please refer to Appendix \ref{appendix-disc}.

\subsection{Multi-Resolution Spectrogram Discriminator}
Along with MFBD, we also apply the multi-resolution spectrogram discriminator (MRSD). Simultaneously using both waveform domain discriminator and time-frequency domain discriminator has been proven to enhance the audio quality \cite{soundstream, audiodec}. We obtain complex spectrograms with Fourier transform sizes of $\{128, 256, 512, 1024\}$ and hop lengths of $\{32, 64, 128, 256\}$. We represent each spectrogram as a real-valued tensor with two channels (real and imaginary) and pass it through a discriminator composed of six 2D convolutions. The model architecture is almost the same as that of \cite{encodec}, except that ours has output channels of $\{16, 16, 32, 64, 128, 1\}$ instead of $\{32, 32, 32, 32, 32, 1\}$ used in \cite{encodec}. We found that this modification results in faster training while maintaining similar audio quality.

\section{Experimental Setup}
\subsection{Training Objective}
Let $\mathcal{G}$ denote a generator composed of an encoder $\mathcal{G}_e$, a quantizer $\mathcal{Q}$, and a decoder $\mathcal{G}_d$. Let $\mathcal{D}_1$ denote MFBD, $\mathcal{D}_2$ denote MRSD, $x$ denote the input waveform and $\hat{x}=\mathcal{G}(x)$ denote the waveform reconstructed by the generator.

\subsubsection{Reconstruction Loss}
 To reduce the distortion in $\hat{x}$, we employ a reconstruction loss. It comprises the mean absolute errors and mean squared errors between multi-resolution log mel spectrograms of $x$ and $\hat{x}$ as given by
\begin{equation}
    \mathcal{L}_{rec} = \sum_{i \in \{1, \cdots, 6\}} \| \mathcal{S}_i(x)-\mathcal{S}_i(\hat{x})\|_1+\| \mathcal{S}_i(x)-\mathcal{S}_i(\hat{x})\|_2^2,
\end{equation}
where $\|\cdot\|_1$ represents a mean of absolute values of elements in a given tensor, $\|\cdot\|_2^2$ represents a mean of squared values, and $\mathcal{S}_i$ represents a log mel spectrogram computed with the Fourier transform size of $2^{i+4}$, hop length of $2^{i+2}$ and the number of mel filters set to the $i$-th element of the set \{6, 12, 23, 45, 88, 128\}. The number of mel filters is determined such that every mel filter has at least one non-zero value \cite{descript}.

\subsubsection{GAN Loss}
We adopt the GAN loss in \cite{encodec} where the generator loss is defined as
\begin{equation}
    \mathcal{L}_{g} = \sum_{j}  \mathbb{E} \Big[ \max \big(0, 1 - \mathcal{D}_j(\hat{x}) \big) \Big],
\end{equation}
and the discriminator loss is defined as
\begin{equation}
    \mathcal{L}_{d}=\sum_{j}\mathbb{E}\left[\max\big(0,1-\mathcal{D}_j(x)\big) + \max\big(0,1+\mathcal{D}_j(\hat{x})\big)\right].
\end{equation}
 
\subsubsection{Feature Matching Loss}
We use the normalized feature matching loss \cite{encodec} as
\begin{equation}
    \mathcal{L}_{fm} = \sum_{j}\sum_{l}\Bigg[\frac{\|\mathcal{D}_j^l(x) - \mathcal{D}^l_j(\hat{x})\|_1}{\|\mathcal{D}_j^l(x)\|_1} \Bigg],
\end{equation}
where $\mathcal{D}_j^l$ is the $l$-th layer of the $j$-th discriminator

\subsubsection{Commitment Loss}
To further reduce the quantization noise, we impose the commitment loss \cite{VQVAE} to the encoder $\mathcal{G}_e$ as given by
\begin{equation}
    \mathcal{L}_c = \left\Vert
        \mathcal{G}_e(x) - \mathcal{Q} \Big(
            \text{sg} \big[ \mathcal{G}_e(x) \big]
        \Big)
    \right\Vert_2^2,
\end{equation}
where sg[$\cdot$] stands for the stop-gradient operator. Since the codebooks in the residual vector quantizer are updated using an exponential moving average of $\mathcal{G}_e(x)$, we do not give explicit loss to the codebooks.

\subsubsection{Loss Balancer}
While the training objective for discriminators is solely composed of $\mathcal{L}_d$, the training objective for the generator is a combination of $\mathcal{L}_g$, $\mathcal{L}_{fm}$, $\mathcal{L}_{rec}$ and $\mathcal{L}_c$. Naturally, this requires an extensive hyperparameter search to find an optimal coefficient for each loss term. To address this issue, we adopted a loss balancer introduced in \cite{encodec} because of its robustness to loss coefficients and ability to stabilize training. Without a loss balancer, $\mathcal{G}$ is trained with a loss
\begin{equation}
    \mathcal{L} = \lambda_{rec}\cdot\mathcal{L}_{rec} + \lambda_{g}\cdot\mathcal{L}_{g} +
                  \lambda_{fm}\cdot\mathcal{L}_{fm}
                = \sum_{i}\lambda_i\cdot\mathcal{L}_i,
\end{equation}
where we temporarily exclude $\mathcal{L}_c$. Then the parameter $w$ of the generator is updated using
\begin{equation}
\begin{split}
    \nabla_w
    &:=\frac{\partial\mathcal{L}}{\partial w}
    = \frac{\partial\mathcal{L}}{\partial \hat{x}} \frac{\partial\hat{x}}{\partial w}
    = \left( \sum_i \lambda_i \cdot \frac{\partial \mathcal{L}_i}{\partial \hat{x}} \right) \frac{\partial \hat{x}}{\partial w} \\
    &= \left( \sum_i \lambda_i \cdot g_i \right) \frac{\partial \hat{x}}{\partial w},
\end{split}
\end{equation}
where $g_i:=\partial \mathcal{L}_i / \partial \hat{x}$. The loss balancer aims to remove the effect of the size of gradient $\|g_i\|_2$ and retain only the direction $g_i / \|g_i\|_2$ so that $\lambda_i$ becomes the true scale of each loss $\mathcal{L}_i$. Using the loss balancer, $w$ is updated using
\begin{equation}
    \widetilde{\nabla}_w=\left( \sum_i\lambda_i\cdot\frac{g_i}{\mathbb{E}\big[\|g_i\|_2\big]} \right)\frac{\partial\hat{x}}{\partial w} + \lambda_c\cdot\frac{\partial\mathcal{L}_c}{\partial w},
\end{equation}
where $\mathbb{E}[\cdot]$ is an exponential moving average. Since we cannot calculate the gradient of $\hat{x}$ with respect to $\mathcal{L}_c$, we did not apply the loss balancer to $\mathcal{L}_c$.

\subsection{Datasets and Training Settings}
\label{sec:dataset-and-training}
We trained two versions of \textit{HILCodec}: one trained on clean speech, and the other trained on general audio, including clean speech, noisy speech and music. For clean speech datasets, we used speech segments from DNS Challenge 4 \cite{dns4} and the VCTK-0.92 corpus \cite{vctk}. For noisy speech, we mixed clean speech segments with noise segments from DNS Challenge 4 during training on-the-fly. For music, we used Jamendo\footnote{It is distributed in 44.1kHz MP3 format. MP3 compression often discards high-frequency components. However, we downsampled the training dataset to 24kHz. Therefore, we suspect that using the training dataset in MP3 format is not problematic.} \cite{jamendo}. All audio files were downsampled to a sampling rate of 24kHz. The total amount of training data is 2,548 hours for clean speech, 175 hours for noise, and 3,730 hours for music. During training on general audio, we sampled each type of audio with the same probability. We randomly extracted a one-second-long segment from each audio file, applied a random gain in the range of [-10dB, 6dB], and scaled down the audio when a clipping occurred during a mixing or a random gain adjustment. 

For GAN training, we adopted simultaneous GAN \cite{simul-gan}. Each model was trained for 468k iterations using 2 RTX 3090 GPUs with a batch size of 24 per GPU, which took 114 hours. We used the AdamP optimizer \cite{adamp} with an initial learning rate of $5\cdot10^{-4}$ and a weight decay of $10^{-5}$. We utilized a cosine annealing learning rate scheduler \cite{coslr} with a linear warmup for 5k iterations.

\subsection{Baselines}
We compare \textit{HILCodec} to various end-to-end neural audio codecs that are publicly available. EnCodec \cite{encodec} and Descript Audio Codec \cite{descript} are models trained on general audio. HiFi-Codec \cite{hificodec} and AudioDec \cite{audiodec} are models trained on clean speech. For all models, we used versions with a sampling rate of 24kHz. We also included EVS \cite{evs}, the signal processing-based audio codec which has demonstrated state-of-the-art performance among traditional codecs \cite{evs-sub}. It is a hybrid codec that uses LPC when compressing speech and uses transform coding when compressing the others. Since it does not support the sampling rate of 24kHz, we upsampled input audios to 32kHz, compressed with EVS, and downsampled back to 24kHz.

\subsection{Evaluation Metrics}
\subsubsection{Subjective Metric}
To compare \textit{HILCodec} with the baselines, we conducted MUSHRA listening tests \cite{mushra} using webMUSHRA framework \cite{webmushra}. As reference audios, we selected four clean speeches, four noisy speeches, and four music pieces for general audio coding, and 12 clean speeches for speech coding. For each test item, we included a low anchor, which was a low-pass filtered version of the reference audio with a cut-off frequency of 3.5kHz, but excluded a mid anchor with a cut-off frequency of 7kHz. Post-screening was performed to exclude assessors who rated hidden references lower than a score of 90 more than 15\% of the test items or rated low anchors higher than 90 more than 15\% of the test items. The number of assessors after post-screening was 11.

\subsubsection{Objective Metric}
\label{sec:settings-eval-obj}
Conducting listening tests is the most reliable way to measure audio quality, but it is time-consuming and costly. Therefore, for model developments and ablation studies, we utilized ViSQOL Audio v3 \cite{visqol}. We selected 100 audio files at a sampling rate of 24kHz for each of the three categories, upsampled reference and coded audio files to a sampling rate of 48kHz, and calculated ViSQOL scores. Further objective evaluations are provided in Appendix \ref{sec:appendix-obj}.

\section{Results}
\begin{figure*}
    \centering
    \includegraphics[width=1.0\linewidth]{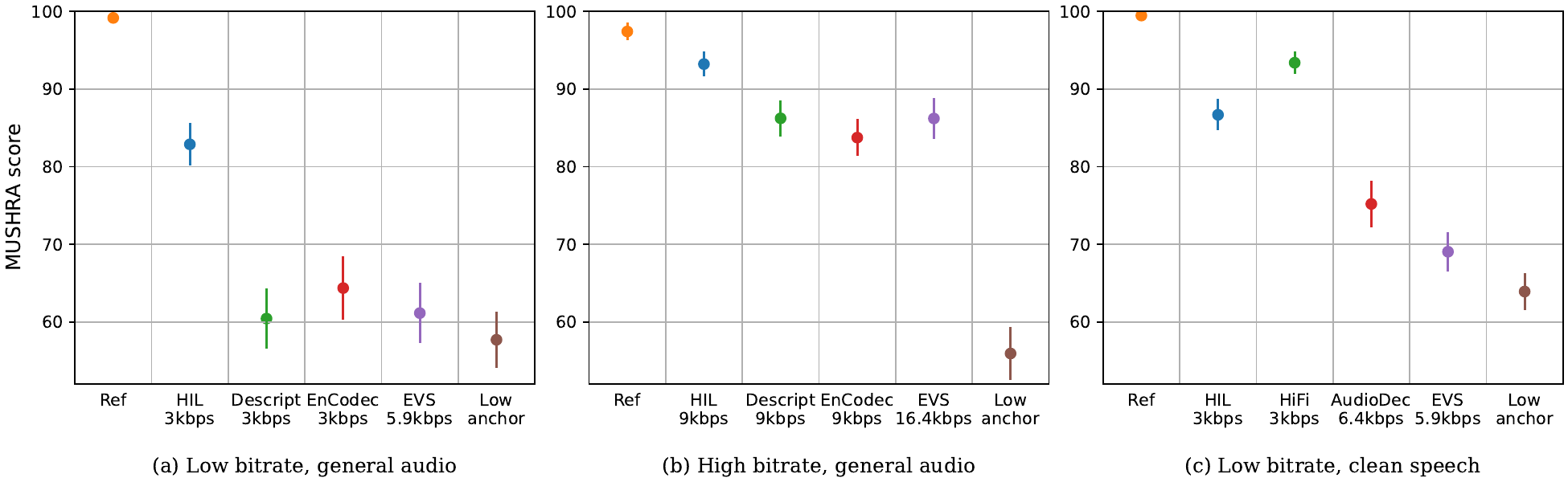}
    \caption{MUSHRA scores of various codecs with 95\% confidence intervals. (a) Low bitrate, trained and evaluated on general audio. (b) High bitrate, trained and evaluated on general audio. (c) Low bitrate, trained and evaluated on clean speech.}
    \label{fig:mushra}
\end{figure*}

\begin{figure*}
    \centerline{\includegraphics[width=1.0\linewidth]{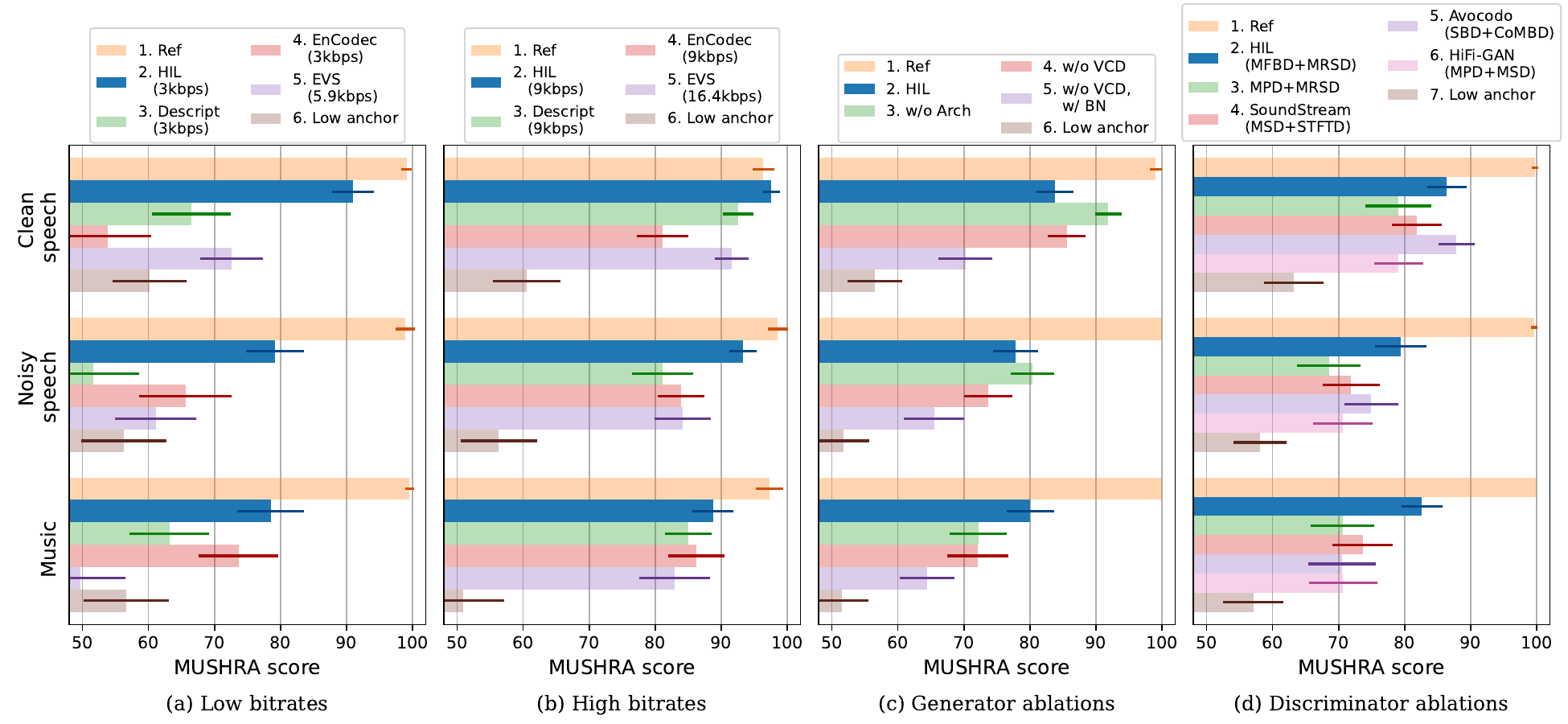}}
    \caption{MUSHRA scores for different audio types with 95\% confidence intervals. (d) "w/o Arch" denotes \textit{HILCodec} without $L_2$-normalization and spectrogram blocks, "w/o VCD" denotes \textit{HILCodec} without the variance-constrained design, and "w/o VCD, w/ BN" denotes \textit{HILCodec} without the variance-constrained design and with batch normalization layers.}
    \label{fig:mushra-atype}
\end{figure*}

\subsection{Subjective Quality Comparison}
Fig. \ref{fig:mushra} illustrates the MUSHRA scores obtained from \textit{HILCodec} and the other baseline codecs. Note that \textit{HILCodec}, EnCodec, AudioDec, and EVS are streamable while Descript and HiFi-Codec are not. Also, \textit{HILCodec} has a complexity smaller than most of the other neural codecs (Section \ref{sec:complexity}). In Fig. \ref{fig:mushra}(a), for general audio coding, \textit{HILCodec} at 3kbps significantly outperforms Descript at 3kbps, EnCodec at 3kbps, and EVS at 5.9kbps. Moreover, in Fig. \ref{fig:mushra}(b), \textit{HILCodec} at 9kbps surpasses other neural audio codecs at 9kbps and EVS at 16.4kbps. In Fig. \ref{fig:mushra}(c), for clean speech coding, \textit{HILCodec} at 3kbps has slightly lower sound quality than high-complexity unstreamable HiFi-Codec at 3kbps. However, it outperforms AudioDec at 6.4kbps and EVS at 5.9kbps by significant margins. Fig. \ref{fig:mushra-atype}(a) and (b) show MUSHRA scores for different audio types. \textit{HILCodec} outperforms other codecs across all bitrates and audio types.

\subsection{Ablation Studies}
\label{sec:ablation}

We conducted ablation studies to prove the effectiveness of the proposed techniques. Due to training resource limits, all models presented in this subsection were trained with a smaller timestep of 156k iterations. All models were trained on a general audio corpus and evaluated at 3kbps.

\subsubsection{Generator}
\label{sec:ablation-generator}

\begin{table}
\caption{ViSQOL scores for ablation studies of the generator. Mean and 95\% confidence interval values are reported.}
\label{table:gen-visqol}
\small
\centering
\setlength{\tabcolsep}{5pt}
\renewcommand{\arraystretch}{1.2}
\begin{tabular}{ p{15em} | C{2cm} }
\hline
Generator ablation & ViSQOL \\
\hline
\textit{HILCodec} & \textbf{4.105 {\footnotesize $\pm$ 0.024}} \\
\hline
\textbf{Architecture} & \\
\quad No $L_2$-normalization & 3.904 {\footnotesize $\pm$ 0.025} \\
\quad No spectrogram blocks & 4.081 {\footnotesize $\pm$ 0.026} \\
\hline
\textbf{Variance-Constrained Design} & \\
\quad Plain residual blocks & 4.082 {\footnotesize $\pm$ 0.026} \\
\quad No input\slash output normalization & 4.059 {\footnotesize $\pm$ 0.024} \\
\quad No zero initialization & 4.092 {\footnotesize $\pm$ 0.023} \\
\hline
\end{tabular}
\end{table}

\begin{table}
\caption{MUSHRA scores for ablation studies of the generator. Mean and 95\% confidence interval values are reported.}
\label{table:gen-mushra}
\small
\centering
\setlength{\tabcolsep}{5pt}
\renewcommand{\arraystretch}{1.2}
\begin{tabular}{ p{15em} | C{2cm} }
\hline
Generator ablation & MUSHRA \\
\hline
\textit{HILCodec} & 80.58 {\footnotesize $\pm$ 1.89} \\
\hline
Without architectural changes & 81.48 {\footnotesize $\pm$ 2.32} \\
Without VCD & 77.15 {\footnotesize $\pm$ 2.35} \\
Without VCD, with BatchNorm & 66.71 {\footnotesize $\pm$ 2.44} \\
\hline
\end{tabular}
\end{table}

\begin{figure}
    \centerline{\includegraphics[width=\columnwidth]{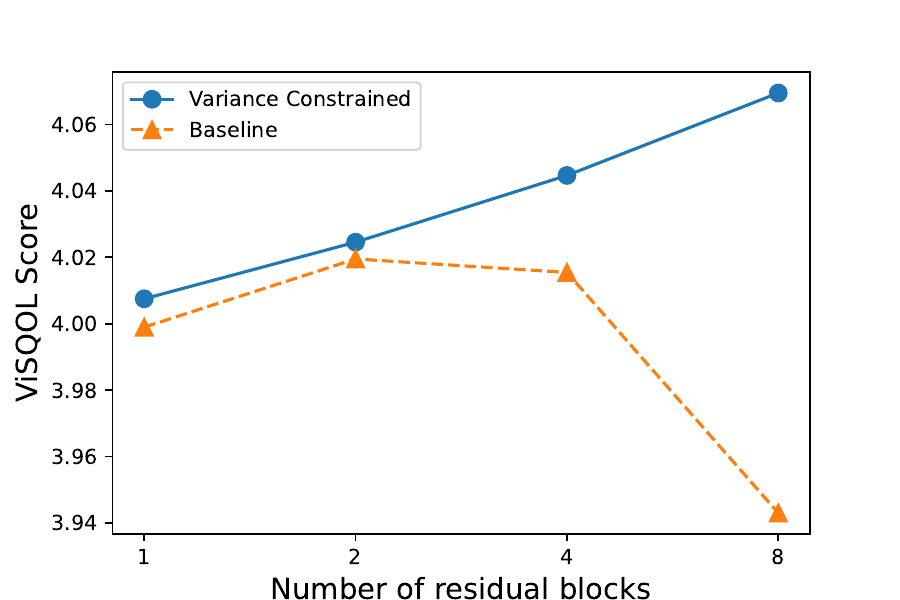}}
    \caption{ViSQOL scores according to the number of residual blocks in each encoder and decoder block. We train each model three times with different seeds and plot the mean value.}
    \label{fig:visqol}
\end{figure}

Starting from \textit{HILCodec}, we removed each proposed generator component while maintaining the other components and measured ViSQOL scores. The results are presented in Table \ref{table:gen-visqol}. Regarding the architectural changes, removing $L_2$-normalization and spectrogram blocks each results in quality degradation. Moreover, replacing the variance-constrained residual blocks with plain residual blocks, removing input/output normalization, and removing zero initialization each results in a lower ViSQOL score.

We conducted a MUSHRA test for the ablation study. Comparing too many models at once increases the difficulty of the MUSHRA test and the burden on assessors. Therefore, we compared four models as follows - \textit{HILCodec}, \textit{HILCodec} without the proposed architectural changes, \textit{HILCodec} without VCD, and \textit{HILCodec} without VCD and with Batch Normalization. The MUSHRA scores are presented in Table \ref{table:gen-mushra}, and the MUSHRA scores for different audio types are presented in Fig. \ref{fig:mushra-atype}(c). It is important to note that mixed-precision training failed for all models except the baseline \textit{HILCodec}. We made three observations. Firstly, in Table \ref{table:gen-mushra}, the difference in MUSHRA scores between \textit{HILCodec} and \textit{HILCodec} without architectural changes is within the margin of error. Upon closer examination, the proposed architectural changes degrade the clean speech quality while improving the music quality. We suspect this is a characteristic of $L_2$-normalization. Practitioners may choose whether to use $L_2$-normalization depending on their audio type of interest. However, one should keep in mind that discarding $L_2$-normalization resulted in the failure of mixed-precision training. Secondly, when VCD is removed, the MUSHRA score degrades significantly. The audio quality remains similar for clean speech, slightly degrades for noisy speech, and significantly degrades for music. This validates the beneficial effect of VCD on the audio quality. Lastly, using Batch Normalization instead of VCD definitely degrades the subjective quality. This implies that Batch Normalization is ill-suited for the SEANet architecture, providing the motivation for VCD (Section \ref{sec:problem-of-seanet}).

To further analyze the impact of variance-constrained design, we performed additional experiments as follows. We built a smaller version of \textit{HILCodec} with $C_{enc}=32$ and $C_{dec}=32$, which we call the variance-constrained model. We also built a baseline model that was exactly the same as the variance-constrained model, except that it did not follow the proposed variance-constrained design: it consisted of plain residual blocks, discarded input/output normalization, and did not employ zero initialization. We tracked the ViSQOL scores at 3kbps while varying the number of residual blocks in each encoder and decoder block from 1 to 8. We trained each model with three different seeds and computed the mean values. In Fig. \ref{fig:visqol}, the objective quality of the baseline model starts to degrade after the number of residual blocks becomes four. On the other hand, the variance-constrained model shows steadily increasing ViSQOL scores along the network depth. This demonstrates the validity of the variance-constrained design.

\subsubsection{Discriminator}
\label{sec:ablation-disc}

\begin{table}
\caption{MUSHRA scores of different discriminators. Mean and 95\% confidence interval values are reported.}
\label{table:disc}
\small
\centering
\setlength{\tabcolsep}{5pt}
\renewcommand{\arraystretch}{1.2}
\begin{tabular}{ p{15em} | C{2cm} }
\hline
Discriminators & MUSHRA \\
\hline
MFBD+MRSD (\textit{HILCodec}) & \textbf{82.81 {\footnotesize $\pm$ 1.96}}\\
MPD+MRSD & 72.72 {\footnotesize $\pm$ 2.84}\\
MSD+STFTD (SoundStream) & 75.82 {\footnotesize $\pm$ 2.50}\\
CoMBD+SBD (Avocodo) & 77.77 {\footnotesize $\pm$ 2.62}\\
MPD+MSD (HiFi-GAN) & 73.47 {\footnotesize $\pm$ 2.62}\\
\hline
\end{tabular}
\end{table}

To demonstrate the effectiveness of the proposed discriminators (MFBD and MRSD), we compared the subjective qualities of various discriminators while maintaining the same generator and training configurations. We included various combinations of discriminators from previously successful neural vocoders and neural audio codecs: MPD and MSD from HiFi-GAN, CoMBD and SBD from Avocodo, and MSD and a STFT discriminator (STFTD) from SoundStream. We also included the combination of MPD and MRSD to validate the superiority of MFBD. Since CoMBD operates on multiple resolutions, we modified the generator for the combination of CoMBD and SBD so that the last three decoder blocks generate waveforms with different sampling rates, as in Avocodo. This modification made it impossible to use the loss balancer. Instead, we strictly followed the loss functions and loss coefficients used in \cite{avocodo}.

Table \ref{table:disc} presents the MUSHRA scores for different combinations of discriminators. The combination of MFBD and MRSD offers the best perceptual quality. Moreover, replacing the proposed MFBD with MPD reduces the subjective quality. Given that the only difference between MFBD and MPD is the use of a filter bank before downsampling, this result demonstrates that our simple solution leads to a considerable improvement in audio quality.

Fig. \ref{fig:mushra-atype}(d) illustrates the MUSHRA scores for different audio types. For clean speech, our discriminators and Avocodo's discriminators both show the best quality, and the difference in scores between the two is within the margin of error. For noisy speech, our disrciminators show the best quality. For music, our discriminators outperform the others by large margins. These results demonstrate that the proposed discriminators are effective across various audio types.

\subsection{Computational Complexity}
\label{sec:complexity}

\begin{table}
\caption{Complexity of various neural audio codecs.}
\label{table:complexity}
\small
\centering
\setlength{\tabcolsep}{3pt}
\renewcommand{\arraystretch}{1.2}
\begin{tabular}{@{ \extracolsep{8pt}}c c c c c c  @{ }} %{ p{15em} | C{2cm} }
\hline
\multirow{2}{*}{Model} & \multirow{2}{*}{\specialcell{\#Params\\(M)}} & \multicolumn{2}{c}{MAC (G)} & \multicolumn{2}{c}{RTF} \\
\cline{3-4}\cline{5-6}
& & Enc & Dec & Enc & Dec \\
\hline
\textit{HILCodec} & 9.58 & 3.29 & 7.46 & 2.47 & 1.13 \\
EnCodec & 14.85 & 1.65 & 1.49 & 4.22 & 4.25 \\
AudioDec & 23.27 & 3.35 & 23.89 & 2.96 & 0.94 \\
Descript & 74.18 & 3.41 & 30.23 & \multicolumn{2}{c}{-} \\
HiFi-Codec & 63.63 & 20.86 & 21.15 & \multicolumn{2}{c}{-} \\
\hline
\end{tabular}
\end{table}

We compared \textit{HILCodec} against other neural audio codecs in terms of computational complexity. We measured the number of parameters, multiply-add counts (MAC) for coding one second long input waveform, and real-time factor (RTF). RTF is defined as the ratio of the length of the input waveform to the time consumed for coding. An RTF higher than 1 means the codec operates in real-time. When calculating RTF, we used one thread of a server CPU (Intel Xeon 6248R 3GHz), exported each model to ONNX Runtime \cite{onnx}, and simulated a streaming condition. Namely, for \textit{HILCodec} and EnCodec, a chunk with 320 waveform samples was encoded and decoded at a time, and for AudioDec, a chunk with 300 samples was processed at a time. For every convolutional layer with a kernel size larger than one, previous samples were stored in a cache and used to calculate the current chunk. Descript and HiFi-Codec are not streamable, so their RTFs are not reported. The results are shown in Table \ref{table:complexity}. \textit{HILCodec} has the smallest number of parameters, requiring minimum memory to store and load model parameters. Moreover, the MAC of \textit{HILCodec} is smaller than AudioDec, Descript, and HiFi-Codec, and its RTF is higher than 1.

\begin{figure*}
    \centering
    \includegraphics[width=1.0\linewidth]{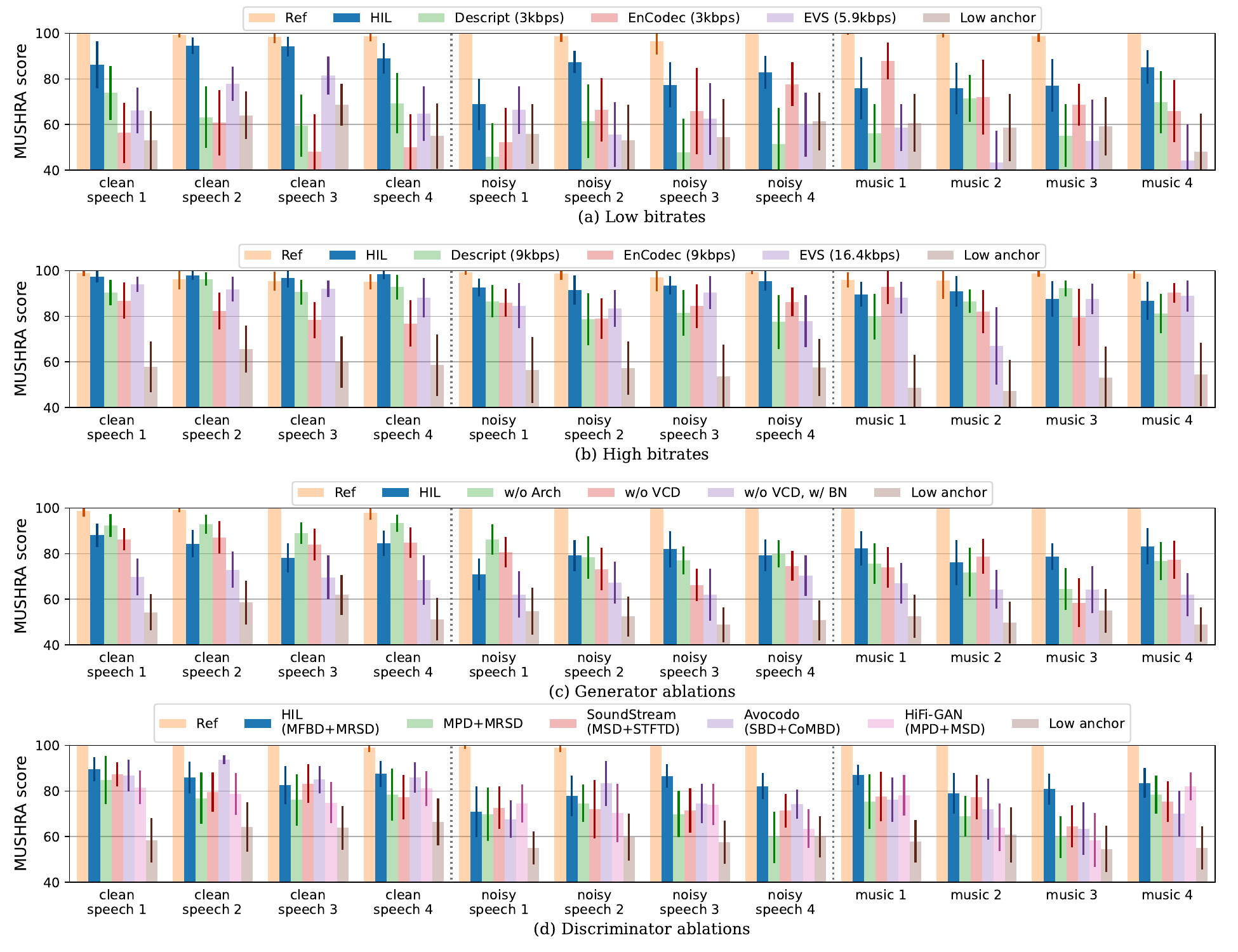}
    \caption{MUSHRA scores with 95\% confidence intervals for all audio items.}
    \label{fig:mushra-all}
\end{figure*}

\section{Conclusion}
We have addressed the issues of previous neural audio codecs: huge complexity, exponential variance growth of the generator, and distortions induced by the discriminator. As a solution, we have proposed a lightweight backbone architecture, a variance-constrained design, and a distortion-free discriminator using multiple filter banks. Combining these elements, the proposed codec, \textit{HILCodec}, achieves superior quality compared to other deep learning-based and signal processing-based audio codecs across various bitrates and audio types. Although we concentrate on building a real-time streaming codec in this work, we expect that our findings can be applied to construct larger and more powerful models.

\appendices
\section{Objective Evaluations}
\label{sec:appendix-obj}
Although we have already conducted the MUSHRA listening test, we are sharing the objective evaluations as they may be beneficial to the research community. We measured ViSQOL Audio v3 \cite{visqol}, PEAQ \cite{peaq}, and the 2f-model \cite{2f-model} for different codecs at various bitrates. PEAQ ranges from $-4$ to $0$, and the 2f-model ranges from $0$ to $100$, with higher scores indicating better quality for both. For PEAQ, we calculated Objective Difference Grade (ODG) using the open-source tool PQevalAudio\footnote{http://www-mmsp.ece.mcgill.ca/Documents/Software/} \cite{pqevalaudio}. We calculated the 2f-model scores using a parameter set\footnote{https://www.audiolabs-erlangen.de/resources/2019-WASPAA-SEBASS\#NewModelParams} calibrated for PQevalAudio as given by
\begin{multline}
    \text{Score} = \frac{56.1345}{1 + (-0.0282 \times \text{AvgModDiff1} - 0.8628)^2} \\
    - 27.1451 \times \text{ADB} + 86.3515,
\end{multline}
\begin{equation}
    \text{FinalScore} = \max(\min(\text{Score}, 100), 0),
\end{equation}
where AvgModDiff1 and ADB are model output variables from PQevalAudio. As we mentioned in the Section \ref{sec:settings-eval-obj}, we used 300 input audio items at a sampling rate of 24kHz for objective evaluations. All the input and the coded output audios are upsampled to a sampling rate of 48kHz before calculating objective metrics. Since EVS doesn't support a sampling rate of 24kHz, we upsampled the input audio to 32kHz, compressed with EVS, and upsampled to 48kHz. The results are presented in Table \ref{table:obj}.

\begin{table}
\caption{Objective And Subjective Evaluations}
\label{table:obj}
\small
\centering
\setlength{\tabcolsep}{5pt}
\renewcommand{\arraystretch}{1.2}
\begin{tabular}{ c c c c c }
\hline
Codec (kbps) & ViSQOL & PEAQ & 2f-model & MUSHRA \\
\hline
\textit{HILCodec} (3) & 4.128 & -3.775 & \textbf{42.897} & \textbf{82.879} \\
Descript (3) & \textbf{4.149} & \textbf{-3.774} & 42.431 & 60.439 \\
EnCodec (3) & 4.117 & -3.804 & 41.616 & 64.364 \\
EVS (5.9) & 2.893 & -3.805 & 31.218 & 61.144 \\
\hline
\textit{HILCodec} (9) & 4.407 & -3.502 & \textbf{56.282} & \textbf{93.212} \\
Descript (9) & \textbf{4.414} & \textbf{-3.454} & 55.743 & 86.277 \\
EnCodec (9) & 4.332 & -3.665 & 50.739 & 83.742 \\
EVS (16.4) & 4.194 & -3.663 & 42.819 & 86.205 \\
\hline
\end{tabular}
\end{table}

\section{MUSHRA Scores For All Items}
Since there are only 12 audio items for each of our MUSHRA evaluations, we present MUSHRA scores for all the 12 items in Fig. \ref{fig:mushra-all}. Note that \textit{HILCodec} evaluated in (a) and (b) is trained for 468k iterations, while \textit{HILCodec} evaluated in (c) and (d) is trained for 156k iterations. Some reference items have no confidence interval, indicating unanimous agreement among all assessors on a score of 100.

\section{Variance of Residual Branches}
\label{appendix-variance}
In Section \ref{sec:VCRB}, we construct the VCRB such that the variance after initialization follows \eqref{resunit-var}. Nevertheless, in Fig. \ref{fig:avg-ch-var}, although the variance increases linearly with the network depth, it does not strictly follow \eqref{resunit-var}. It is demonstrated that for exactness, we must use scaled weight standardization (Scaled WS) \cite{nfnet1}. However, in our early experiments, using Scaled WS resulted in a slightly worse ViSQOL score. Reference \cite{nfnet1} also experienced similar degradation and hypothesized that Scaled WS harms the performance of depthwise convolutions. Therefore, as an alternative, we carefully apply initialization for each layer. For layers where an activation function does not come immediately after, we apply LeCun initialization so that $\mathbf{Var}(W\mathbf{x}) \approx \mathbf{Var}(\mathbf{x})$ where $W$ is a weight parameter. For layers with subsequent activation functions, we apply He initialization so that $\mathbf{Var}(\sigma (W\mathbf{x}))\approx \mathbf{Var}(\mathbf{x})$. Through these methods, despite the model does not exactly follow \eqref{resunit-var}, it achieves a linear variance increment.

\section{Differences Between MFBD and MSD}
\label{appendix-disc}
We present detailed comparisons between SBD used in Avocodo and MFBD used in \textit{HILCodec}. First, SBD decomposes the input signal into 16 sub-bands while MFBD uses co-prime numbers of sub-bands $N \in \{1,2,3,5,7,11\}$. As described in Section \ref{sec:mfbd}, this enables MFBD to model both long- and short-term dependencies and handle distortions in transition bands. Second, in SBD, a single sub-discriminator accepts the entire 16 sub-bands simultaneously using a convolutional layer with input channels 16. On the other hand, in MFBD, a sub-discriminator accepts each sub-band separately using a convolutional layer with input channels of 1. However, the parameter of the sub-discriminator is shared across sub-bands as illustrated in Fig. \ref{fig:disc}(c). Third, in SBD, there exists a sub-discriminator that operates on a transposed sub-band feature. This means that the number of sub-bands becomes the length of the input, and the length of the sub-band feature becomes the channel size of the input. In MFBD, we do not employ such a sub-discriminator. We suspect these overall differences result in the superior performance of MFBD over SBD.

\section{Selection of Audio Samples for MUSHRA}
\label{appendix-audio}
We provide our considerations for selecting audio samples for the MUSHRA experiments. First, we had to choose which audio domains to evaluate. Upon reviewing previous papers, we found that the domains evaluated varied slightly. SoundStream \cite{soundstream} utilized clean speech, noisy speech, noisy and reverberant speech, and music. EnCodec \cite{encodec} employed clean speech, noisy speech, music, and proprietary music. Descript Audio Codec \cite{descript} used clean speech, environmental sounds, and music. We chose to evaluate the most commonly appearing domains: clean speech, noisy speech, and music. Second, we tried to include various sounds. For noisy speech, we included keyboard typing noise, fan noise, the sound of opening chip packets, and the sound of water flowing. The genres of the music samples were pop, soul, electronic, and country rock. Third, for speech codecs, we selected challenging audio samples. We coded audio candidates using an early version of \textit{HILCodec}, and selected those with large distortions. In our experience, this selection was crucial because the majority of neural speech codecs coded randomly selected speech samples very well, making those samples unsuitable for comparing each model.

\section*{References}

\def\refname{}

\end{document}